\documentclass{article}

\usepackage{graphicx}
\begin{document}

\title{Transmission of Renormalized Benzene Circuits}

\date{May 13, 2015}

\author{Kenneth W. Sulston\thanks{corresponding author} \thanks{Department of Mathematics and Statistics, University of Prince Edward Island, Charlottetown, PE, C1A 4P3, Canada}\and Sydney G. Davison\thanks{Department of Applied Mathematics, University of Waterloo, Waterloo, ON, N2L 3G1, Canada} \thanks{Department of Physics and the Guelph-Waterloo Physics Institute, University of Waterloo Campus, Waterloo, ON, N2L 3G1, Canada}}

\maketitle

\section{Abstract}

The renormalization equations emerge from a Greenian-matrix solution of the discretized Schr\"{o}dinger equation.
A by-product of these equations is the decimation process, which enables substituted-benzenes to be mapped onto
corresponding dimers, that are used to construct the series and parallel circuits of single-, double- and triple-dimers.
The transmittivities of these circuits are calculated by the Lippmann-Schwinger theory, which yields the transmission-energy function $T(E)$. The average value of $T(E)$ provides a measure of the electron transport in the circuit in question. The undulating nature of the $T(E)$ profiles give rise to resonances ($T=1$) and anti-resonances ($T=0$) across the energy spectrum.  Analysis of the structure of the $T(E)$ graphs highlights the distinguishing features associated with the homo- and hetero-geneous series and parallel circuits. Noteworthy results include the preponderance of p-dimers in circuits with high $T(E)$ values, and the fact that parallel circuits tend to be better transmitters than their series counterparts.

\section{Introduction}

Quantum-interference effects play a dominant role in the electron transmission through substituted-benzene molecules, and give rise to markedly different transmission-energy $T(E)$ profiles. These effects occur, because of the phase-shifts introduced into the tunnelling electron wave-function, via the different spatial pathways taken by the electron, during its journey through the molecule.

Sautet and Joachim \cite{ref1} adopted the electron scattering quantum chemistry technique \cite{ref2}, in conjunction with the extended H\"{u}ckel molecular orbital method \cite{ref3}, to demonstrate the existence of such electronic interferences for an electron tunnelling through a single benzene molecule inserted in a polyacetylene chain $(CH)_n$. It was found that the phase-shift was sensitive to the change in the connection conformation between the benzene and the two $(CH)_n$ molecular wires. In the case of para (p)-benzene, the bifurcation pathways were of equal length, so that constructive interference took place, while in the meta (m)- and ortho (o)-benzene situations, the electron pathways were of unequal lengths, whereby destructive interference resulted. Thus, the p-configuration was preferred over the m- and o- cases in the benzene transmission. Such findings were reflected in the $T(E)$ curves, which were discussed, using the symmetry and energy of the benzene molecular orbitals compared to the energy of the $(CH)_n$ band.

More recently, Hansen, Solomon and coworkers \cite{ref4} sought to gain a greater understanding of quantum interference by undertaking a comprehensive analytical treatment of substituted-benzene electron transmission based on a Green-function (GF) approach \cite{ref5}, which incorporated the L\"{o}wdin partitioning technique \cite{ref6}.  The antiresonances, arising in the transmisison $T(E)$ (i.e. $T=0$), were attributed to either multi-path zeroes, created by interfering spatial pathways, or resonance zeroes analogous to zeroes induced by sidechains. Further investigations of quantum interference in transmission through single molecules have also been reported\cite{ref6a, ref6b}, where extensive lists of useful references to the literature can be found. The most recent contribution to the analytical solution of electronic transport through a single benzene molecule has been provided by Dias and Peres \cite{ref6c}, who adopted the lattice Green-function approach. The pedagogical style of this article recommends it to students and researchers less conversant with the Green-function technique.

On turning to the question of electron transport in multiple-benzene structures, it is a great advantage to have a means that enables the corresponding elaborate GF to be made more tractable, while still retaining all the electronic information contained in the original GF.  Such a means is provided by the renormalization method \cite{ref7}, whose application was illustrated by the treatment of the density-of-states distribution \cite{ref8} over a benzene molecule with an arbitrary number and positions of the attached polymer chains.  The effect of two chains, connected in various conformations, was studied and the differences between the p-, m- and o- conformations were identified by contrasting them with the single-chain case.

In this article, having become acquainted with the renormalization approach, in which each of the substituted\footnote{While the word ``substitute'' is generally used in such cases where any molecule is attached (i.e., substituted) to the benzene molecule, in the present paper, it is strictly confined to the situation where the leads are attached to the p-, m-, o- benzene sites.} p-, m- and o- benzenes is represented by a corresponding and equivalent rescaled dimer, we study the transmission properties of such rescaled dimers, fabricated  at the molecular level, in the form of series and parallel circuits, which mimic their classical electric-circuit counterparts. In addition, various other types of these dimer circuits are investigated and their $T(E)$ values obtained via the Lippmann-Schwinger (LS) scattering theory \cite{ref9}. A table of the average $T$-values is compiled, from which the ranges of good and poor conductors can be ascertained.  A selection of particular $T(E)$ curves is also provided and discussed.  Such a renormalized GF-LS study should afford useful insight into the electron transport properties of more complex multi-benzene structures of value in designing molecular electronic devices \cite{ref10, ref11}.

In the next section, we develop the renormalized Greenian-matrix approach, whereby the crucial renormalization equations are derived.

\section{Renormalization Equations}

The GF technique has a well-established reputation as a powerful, versatile and flexible method, and the incorporation of the renormalization procedure \cite{ref7, ref8, ref13, ref13a} introduces a degree of simplification that enhances its potential, when treating the electronic properties of complex molecular structures \cite{ref12}.  In a sense, such an incorporation may be viewed as a natural development of the GF method.

The basic renormalization equations can be derived by means of the Greenian-matrix ($G_{mn}$) version of the discretized Schr\"{o}dinger equation, which takes the elemental form
\begin{equation}
(E - \alpha_m) G_{mn} = \delta_{mn} + \sum_k \beta_{mk} G_{kn} ,
\label{eq1}
\end{equation}
where $\alpha$ ($\beta$) denotes the atomic-site (bond) energy, and with the $m$, $n$, $k$ subscripts ranging from 1 to 3, for the case of a trimer molecule considered here (Figure 1). 
\begin{figure}[htbp]
\includegraphics[width=15cm]{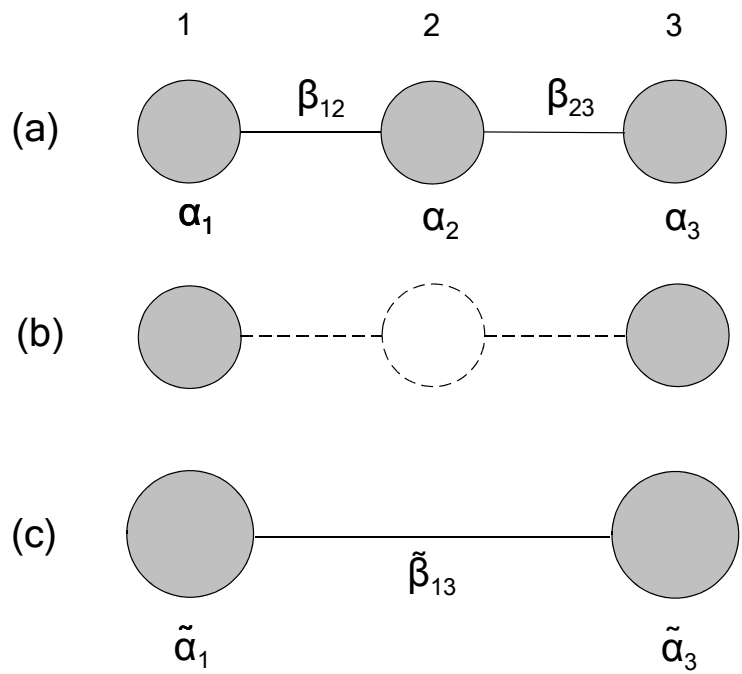}
\caption{Decimation of the site-2 atom of the trimer molecule, resulting in the renormalized dimer molecule.}
\label{fig1}
\end{figure}

Expanding (\ref{eq1}), the complete set of equations for the trimer molecule reads:
\begin{equation}
(E - \alpha_1) G_{11} = 1 +  \beta_{12} G_{21} , ~m=n=1 ,
\label{eq2a}
\end{equation}
\begin{equation}
(E - \alpha_2) G_{22} = 1 + \beta_{21} G_{12}+ \beta_{23} G_{32} , ~m=n=2 ,
\label{eq2b}
\end{equation}
\begin{equation}
(E - \alpha_3) G_{33} = 1 +  \beta_{32} G_{23} , ~m=n=3,
\label{eq2c}
\end{equation}
\begin{equation}
(E - \alpha_1) G_{12} = \beta_{12} G_{22} , ~m=1, n=2 ,
\label{eq2d}
\end{equation}
\begin{equation}
(E - \alpha_1) G_{13} = \beta_{12} G_{23} , ~m=1, n=3 ,
\label{eq2e}
\end{equation}
\begin{equation}
(E - \alpha_2) G_{21} = \beta_{21} G_{11}+ \beta_{23} G_{31} , ~m=2 ,n=1,
\label{eq2f}
\end{equation}
\begin{equation}
(E - \alpha_2) G_{23} = \beta_{21} G_{13}+ \beta_{23} G_{33} , ~m=2 ,n=3,
\label{eq2g}
\end{equation}
\begin{equation}
(E - \alpha_3) G_{31} = \beta_{32} G_{21} , ~m=3, n=1 ,
\label{eq2h}
\end{equation}
\begin{equation}
(E - \alpha_3) G_{32} = \beta_{32} G_{22} , ~m=3, n=2 .
\label{eq2i}
\end{equation}

With the aid of (\ref{eq2f}), equation (\ref{eq2a}) becomes
\begin{equation}
(E - \alpha_1) G_{11} =1+ (\beta_{12} \beta_{21} G_{11}+\beta_{12} \beta_{23} G_{31})
(E-\alpha_2)^{-1}, 
\end{equation}
which can be rewritten as
\begin{equation}
(E - \alpha_1- {{\beta_{12} \beta_{21}}\over {E-\alpha_2}} ) G_{11} =1+  {{\beta_{12} \beta_{23}}\over {E-\alpha_2}}  G_{31},
\end{equation}
or
\begin{equation}
(E - \tilde{\alpha}_1) G_{11} =1+  \tilde{\beta}_{13}   G_{31},
\label{eq3a}
\end{equation}
where
\begin{equation}
\tilde{\alpha}_1 =\alpha_1+ { {\beta}_{12}^2 \over {E-\alpha_2}},
\label{eq4a}
\end{equation}
\begin{equation}
\tilde{\beta}_{13}  = {{\beta_{12} \beta_{23}} \over {E-\alpha_2}}.
\label{eq4b}
\end{equation}
Using (\ref{eq2g}) in (\ref{eq2c}), we find
\begin{equation}
(E - \alpha_3) G_{33} =1+ (\beta_{32} \beta_{21} G_{13}+\beta_{32} \beta_{23} G_{33})
(E-\alpha_2)^{-1}, 
\end{equation}
which on rearranging gives
\begin{equation}
(E - \alpha_3- {{\beta_{32} \beta_{23}}\over {E-\alpha_2}} ) G_{33} =1+  {{\beta_{32} \beta_{21}}\over {E-\alpha_2}}  G_{13}.
\end{equation}
On writing this in the form
\begin{equation}
(E - \tilde{\alpha}_3) G_{33} =1+  \tilde{\beta}_{31}   G_{13},
\label{eq3b}
\end{equation}
we have
\begin{equation}
\tilde{\alpha}_3 =\alpha_3+ { {\beta}_{23}^2 \over {E-\alpha_2}},
\label{eq5a}
\end{equation}
\begin{equation}
\tilde{\beta}_{31}  = {{\beta_{32} \beta_{21}} \over {E-\alpha_2}}.
\label{eq5b}
\end{equation}
Furthermore, equations (\ref{eq4a}), (\ref{eq4b}), (\ref{eq5a}) and (\ref{eq5b}) can also be derived from (\ref{eq2e}) together with (\ref{eq2g}), and (\ref{eq2h}) in conjunction with (\ref{eq2f}).  During the course of these calculations, two other renormalization equations arise, namely,
\begin{equation}
(E - \tilde{\alpha}_1) G_{13} = \tilde{\beta}_{13}   G_{33},
\label{eq5c}
\end{equation}
\begin{equation}
(E - \tilde{\alpha}_3) G_{31} = \tilde{\beta}_{31}   G_{11}.
\label{eq5d}
\end{equation}
Thus, we see that the original set of equations (\ref{eq2a})-(\ref{eq2i}), for the trimer (Figure 1a), has been renormalized into a set of four equations (\ref{eq3a}), (\ref{eq3b}), (\ref{eq5c}) and (\ref{eq5d}), involving only the atomic sites 1 and 3, with site 2 being decimated, so as to form the dimer (Figure 1c).  The rescaled parameters for the dimer are given by equations (\ref{eq4a}), (\ref{eq4b}), (\ref{eq5a}) and (\ref{eq5b}), which lead to the general renormalization equations
\begin{equation}
\tilde{\alpha}_{n-1} =\alpha_{n-1}+ { {\beta}_{n-1,n}^2 \over {E-\alpha_n}},
\label{eq6a}
\end{equation}
\begin{equation}
\tilde{\alpha}_{n+1} =\alpha_{n+1}+ { {\beta}_{n,n+1}^2 \over {E-\alpha_n}},
\label{eq6b}
\end{equation}
\begin{equation}
\tilde{\beta}_{n-1,n+1}  = {{\beta_{n-1,n} \beta_{n,n+1}} \over {E-\alpha_n}}.
\label{eq6c}
\end{equation}
for the effective site energies and intersite coupling of  the renormalized sites $n-1$ and $n+1$, upon decimation of the site $n$.  In practice, the renormalization equations (\ref{eq6a})-(\ref{eq6c}) are applied recursively for the successive elimination of the sites in complex structures \cite{ref12}, until it has been reduced to one or a few sites.  We observe that the renormalization-decimation approach and the L\"{o}wdin matrix-partition scheme \cite{ref6} both contain the rescaling features of equations (\ref{eq6a})-(\ref{eq6c}).

\section{Dimerization of Substituted Benzenes}

The study of electron transmission through a substituted-benzene molecule is facilitated by employing the renormalization technique to reduce the molecule to an equivalent dimer.  The type of dimer involved is governed by the location of the atomic sites to which the molecular leads are connected.  In the p-benzene case, the dimer links the (1,4) sites, while in the m-benzene and o-benzene situations the (1,5) and (1,6) sites are joined, respectively (Figure 2). 
\begin{figure}[htbp]
\includegraphics[width=15cm]{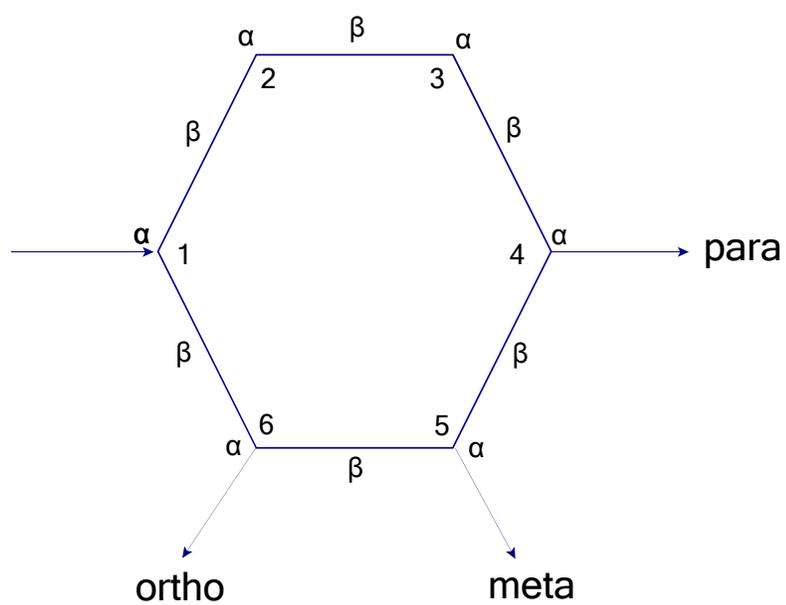}
\caption{Locations of the para-, meta- and ortho-benzene sites where molecular leads are attached to the benzene molecule.}
\label{fig2}
\end{figure}

As we have seen, the renormalization scheme possesses concomitantly a mathematical representation and a diagrammatical visualization of each decimation stage of the molecular reduction.  For brevity, only the latter process will be retained here, together with the expressions for the effective site energies and intersite coupling of the rescaled dimer sites \cite{ref7, ref8, ref13}.

\subsection {Para-benzene dimer}

Starting with the site-6 decimation in Figure 3a, we obtain Figure 3b.  Similarly, we proceed along the site-decimation sequence 5, 2, 3 to arrive at the p-benzene dimer in Figure 3e, whose rescaled relations are
\begin{equation}
\tilde{\alpha}_1 =\check{\alpha}_1+  \bar{\beta}_{13}^2 (E-\bar{\alpha}_3)^{-1},
\label{eq7}
\end{equation}
\begin{equation}
\tilde{\alpha}_4 =\bar{\alpha}_4+  \beta_{34}^2  (E-\bar{\alpha}_3)^{-1},
\label{eq8}
\end{equation}
\begin{equation}
\tilde{\beta}_{14}  = \bar{\beta}_{13} \beta_{34} (E-\bar{\alpha}_3)^{-1},
\label{eq9}
\end{equation}
for the site energies and intersite coupling, respectively.
\begin{figure}[htbp]
\includegraphics[width=14cm]{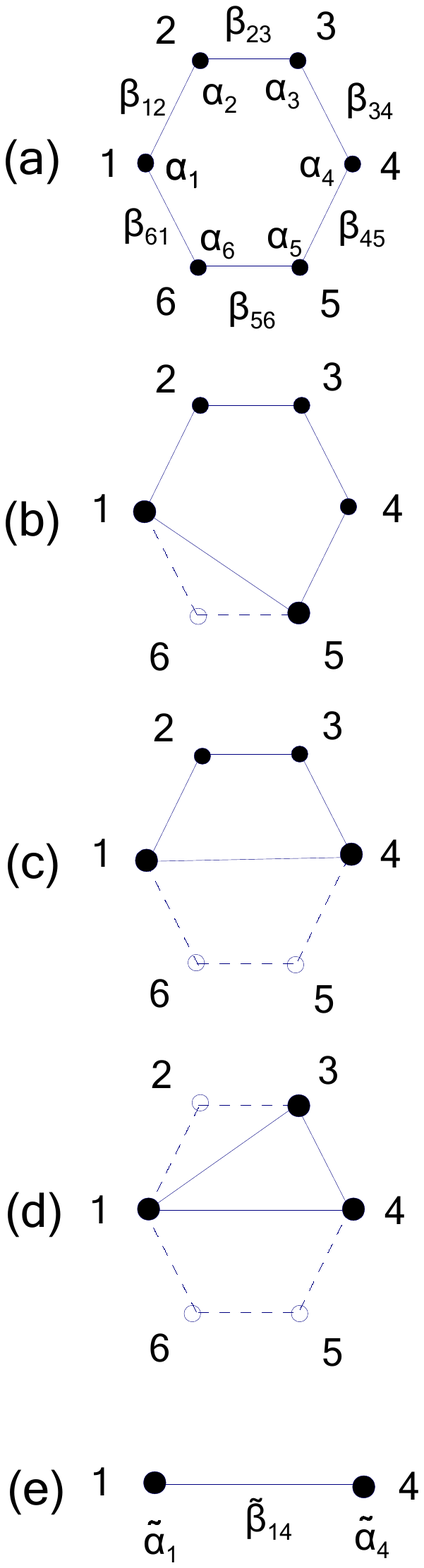}
\caption{Decimation-renormalization scheme for reducing the benzene molecule to a p-benzene dimer. The parts of the system decimated atom-by-atom are shown as dashed lines and open circles, while renormalized sites are represented by slightly enlarged full circles.}
\label{fig3}
\end{figure}

At this juncture, having seen how the renormalization-decimation process maps the original benzene molecule onto the p-benzene dimer (Figure 3), we solve the dimer rescaled equations (\ref{eq7}) to (\ref{eq9}), in conjunction with all the preceding simultaneous equations (not shown), and express the final result in terms of the site ($\alpha$) and bond ($\beta$) energies of the original benzene molecule (Figure 2), whence, we obtain
\begin{equation}
\bar{\alpha}_p \equiv \tilde{\alpha}_1= \tilde{\alpha}_4 = \alpha + \tilde{\beta}_{14} X ,
\label{eq10}
\end{equation}
\begin{equation}
\bar{\beta}_p \equiv \tilde{\beta}_{14} = 2 \beta (X^2-1)^{-1} ,
\label{eq11}
\end{equation}
where we have introduced the dimensionless reduced energy
\begin{equation}
X = (E-\alpha)/\beta .
\label{eq12}
\end{equation}

Adopting the preceding treatment for the p-benzene dimer, the corresponding results for the renormalization of the m- and o-benzene dimers can be summarized, as follows.

\subsection {Meta-benzene dimer}

The successive site-decimations 2, 3, 4 and 6 yield the (1,5) dimer, for which
\begin{equation}
\bar{\alpha}_m \equiv \tilde{\alpha}_1= \tilde{\alpha}_5 = \alpha + \beta X^{-1} + \tilde{\beta}_{15}  ,
\label{eq13}
\end{equation}
\begin{equation}
\bar{\beta}_m \equiv \tilde{\beta}_{15} =  \beta X^{-1} (X^2-1) (X^2-2)^{-1}.
\label{eq14}
\end{equation}

\subsection {Ortho-benzene dimer}

Here, the site-decimations 2, 5, 3 and 4 lead to the (1,6) dimer, with
\begin{equation}
\bar{\alpha}_o \equiv \tilde{\alpha}_1= \tilde{\alpha}_6 = \alpha + \beta (X^2-2) (X^2-X-1)^{-1} - \tilde{\beta}_{16}  ,
\label{eq15}
\end{equation}
\begin{equation}
\bar{\beta}_o \equiv \tilde{\beta}_{16} =  \beta (X^2-1) (X^2-2) [(X^2-1)^2-X^2]^{-1}.
\label{eq16}
\end{equation}

We note that the equality of each of the dimer site-energies reveals that all of the p-, m- and o-dimers are symmetric.  We now turn our attention to the discussion of the circuits consisting of benzene dimers in series and in parallel, and their reduction to the corresponding single dimers, for the purpose of transmission analysis.  As is well-known, in traditional electrical circuits, greater security against transmission breakdown is achieved by connecting resistances in a parallel configuration rather than a series arrangement. Such an advantage should also be beneficial at the quantum level in molecular electronic circuits, where 1-electron transmission is involved.  In the present study, renormalized benzene dimers will play the role of the resistances, while molecular wires will provide the leads to the dimers in series and parallel.

\section{Benzene Dimers in Series}

In mimicking the classical electric circuit of resistances in series, we consider three benzene dimers arranged in such a series (Figure 4), where each dimer can be taken to be any one of the p-, m- or o- dimers, set in any order.  There are $3^3=27$ arrangements, although only 18 of these (9 symmetric and 9 asymmetric) yield unique transmissions, because orderings such as opm and mpo are physically indistinguishable.  Based on the renormalization-decimation recipe, a sequence of four decimations ultimately results in mapping the three dimers in series onto the required single dimer (Figure 4f), which can be symmetric or asymmetric.  We consider the three symmetric dimers to have site (bond) energies $\alpha_n$ ($\beta_n$), where $n=1,2$ or 3.  The remaining inter-dimer bond energies are $\beta$ (Figure 4a).  
\begin{figure}[htbp]
\includegraphics[width=14.5cm]{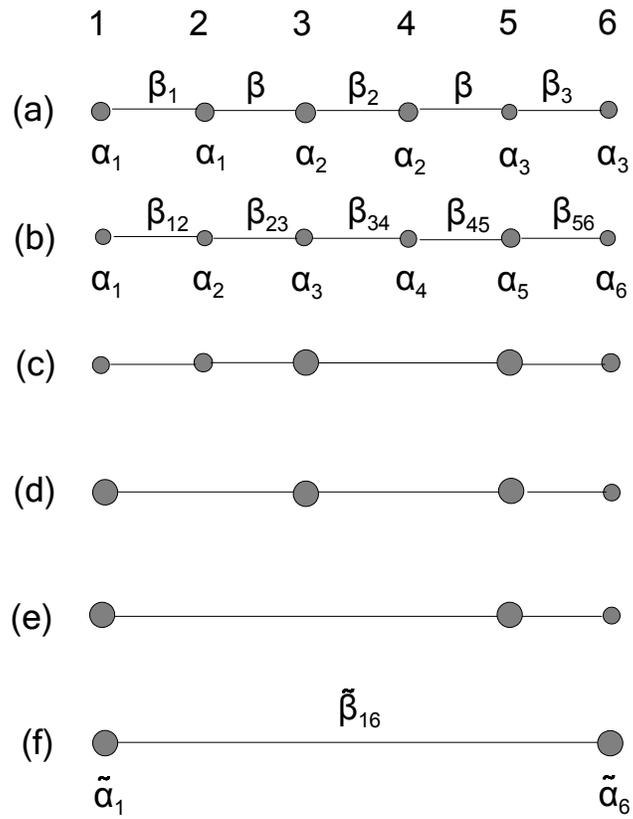}
\caption{Decimation scheme for mapping three symmetric dimers in series onto a single renormalized dimer.}
\label{fig4}
\end{figure}

The renormalization equations, corresponding to successive decimation of the atomic sites in the sequence 4, 2, 3 and 5, can be obtained from the appropriate diagram in Figure 4. Solving all of these equations, we arrive at:
\begin{equation}
\tilde{\alpha}_1 = \alpha_1 + \beta \eta_1^2 X_1^{-1} ( 1+ X_1^{-1} Z_3  \Gamma^{-1}),
\label{eq17}
\end{equation}
\begin{equation}
\tilde{\alpha}_6 = \alpha_3 + \beta \eta_3^2 X_3^{-1} (1+ X_3^{-1} Z_1 \Gamma^{-1}) ,
\label{eq18}
\end{equation}
\begin{equation}
\tilde{\beta}_{16} = \beta \eta_1 \eta_2 \eta_3 X_1^{-1} X_3^{-1} \Gamma^{-1} ,
\label{eq19}
\end{equation}
where
\begin{equation}
X_n=(E-\alpha_n)/\beta ~,~ \eta_n = \beta_n /\beta, ~n=1,2,3,
\label{eq20}
\end{equation}
\begin{equation}
Z_1 =X_2 - X_1^{-1} ~,~ Z_3= X_2 - X_3^{-1},
\label{eq21}
\end{equation}
\begin{equation}
\Gamma = Z_1 Z_3 - \eta_2^2.
\label{eq22}
\end{equation}
We note in (\ref{eq20}) that $(\alpha_n, \beta_n)$ refer to the rescaled parameters of the original isolated p-, m-, o- benzene dimers (found in (\ref{eq10})-(\ref{eq16})).

If dimers 1 and 3 are different, then $\tilde{\alpha}_1  \not= \tilde{\alpha}_6$ in the above equations, and the series dimer is classified as being asymmetric, so its structure is pmo-like.  However, if dimers 1 and 3 are identical, making $\tilde{\alpha}_1 = \tilde{\alpha}_6$, so that the series dimer becomes symmetric, indicating a pmp-like structure, for which (\ref{eq19}) reduces to 
\begin{equation}
\tilde{\beta}_{16} = \beta \eta_1^2 \eta_2  X_1^{-2}  \Gamma^{-1} .
\label{eq23}
\end{equation}

It is worth noting that the results for a double-dimer series can be obtained directly from those of the triple-dimer series by putting $\alpha_3=\alpha$ (at sites 5 and 6) and $\beta_3=\beta$ (between sites 5 and 6), which reduces the third dimer to ``regular'' chain atoms.

\section{Benzene Dimers in Parallel}

In the case of three resistances in parallel, we adopt as a model three symmetric benzene dimers (Figure 5a), to which p-, m- and o- dimers can be assigned in 10 physically distinguishable ways, all of which give rise to a symmetric final dimer.  
\begin{figure}[htbp]
\includegraphics[width=15cm]{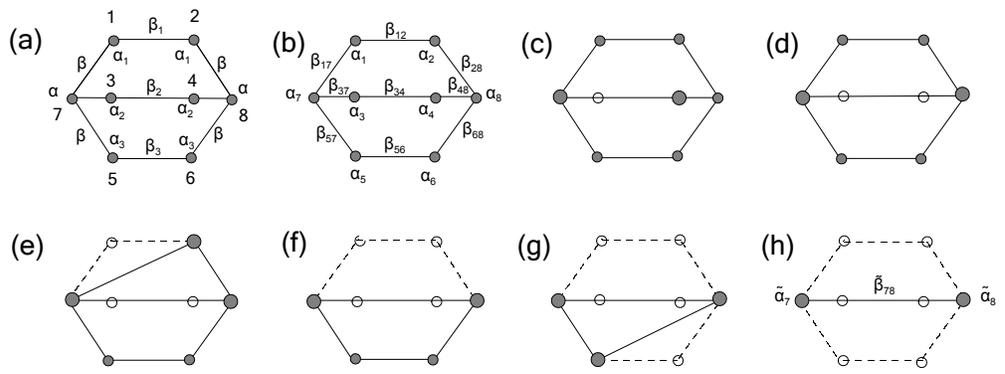}
\caption{Decimation scheme for mapping three symmetric dimers in parallel onto a single renormalized dimer.}
\label{fig5}
\end{figure}
The three dimers are embedded in a parallel structure, and have rescaled parameters $(\alpha_n, \beta_n)$ where $n=1,2,3$. The remaining 6 bonds are labelled $\beta$.  In Figure 5b, we relabel the parameters in the renormalization notation, in preparation for the atomic-site decimation sequence 3, 4, 1, 2, 6, 5 in the subsequent Figures 5c to 5h. Using these figures, we can generate the final renormalization equations
\begin{equation}
\tilde{\alpha}_7 = \check{\alpha}_7 + \beta_{75}^2 (E-\bar{\alpha}_5)^{-1},
\label{eq24}
\end{equation}
\begin{equation}
\tilde{\alpha}_8 = \hat{\alpha}_8 + \bar{\beta}_{58}^2 (E- \bar{\alpha}_5)^{-1},
\label{eq25}
\end{equation}
\begin{equation}
\hat{\beta}_{78} = \beta_{75} \bar{\beta}_{58} (E-\bar{\alpha}_5)^{-1}.
\label{eq26}
\end{equation}
Solving these last three equations by utilizing all the preceding renormalization equations (not shown) for the above decimation sequence, we obtain the expressions:
\begin{equation}
\tilde{\alpha}_7=\tilde{\alpha}_8= \alpha + \beta \sum_{n=1}^3 (X_n - \eta_n)^{-1} - \tilde{\beta}_{78},
\label{eq27}
\end{equation}
\begin{equation}
\tilde{\beta}_{78} = \sum_{n=1}^3 \beta_n (X_n^2 - \eta_n^2)^{-1},
\label{eq28}
\end{equation}
where $X_n$ and $\eta_n$ are given by (\ref{eq20}).  These equations provide the rescaled parameters of the single dimer in Figure 6h, which is the renormalized symmetric dimer of the three symmetric dimers in parallel in Figure 5a.  The corresponding relations for 2 dimers in parallel are obtained by simply letting $n=1..2$ only in the above summations.

\section{Transmittivity of Renormalized Benzene Dimers}

The effects of quantum interference in benzene dimers manifest themselves in their transmission-energy $T(E)$ spectrum profiles.  Such a situation is also encountered in transmission through pairs of impurities \cite{ref14} (adatoms) in (on) atomic wires.  Actually, these latter systems closely resemble one another, and a direct link between them can be established  by means of the renormalization technique.

Let us now address the question of the transmission of a benzene dimer, connected to two semi-infinite atomic-wire leads, as in Figure 6, where $\alpha$ ($\beta$) is the atomic-wire site (bond) energy, and $\alpha_0$, $\alpha_1$ and $\beta_{01}$ are the corresponding dimer energies.  
\begin{figure}[htbp]
\includegraphics[width=15cm]{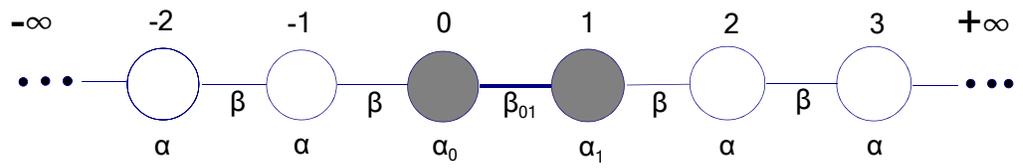}
\caption{Two semi-infinite leads attached to a dimer located between sites 0 and 1.}
\label{fig6}
\end{figure}

Access to the transmittivity of the benzene dimer can be gained via the transmission probability $T(E)$, which is obtained by invoking the LS equation of scattering theory \cite{ref9}, namely,
\begin{equation}
c_n (k) = c_n^0 (k) + \sum_{l, m} G_0(n, l) V (l, m) c_m (k) ,
\label{eq29}
\end{equation}
whose basic formulation appears in Mi\v{s}kovi\'{c} {\em et al} \cite{ref14}.  The Dysonian structure of (\ref{eq29}) contains the $n$-site orbital coefficient $c_n^0 (k)$ of the unperturbed atomic wire, which is represented by Bloch waves of unit amplitude, travelling from left to right, whereby $c_n^0 (k)= e^{i n \theta_k}$, with momentum $\theta_k = ka$, $k$ being the wave number and $a$ the atomic-wire spacing, and reduced energy $\chi_k=(E_k-\alpha)/2 \beta = \cos \theta_k$ lying inside the band ( $|\chi_k| \le 1$).  The Greenian-matrix element \cite{ref16} is given by 
\begin{equation}
G_0(n,m) = \langle n | G_0 | m \rangle = {i e^{i |n-m| \theta_k} \over 2 \beta \sin \theta_k}.
\label{eq30}
\end{equation}
The $n$-site orbital coefficient $c_n(k)$ of the perturbed atomic-wire scattering eigenfunction is represented, far to the left, by an incoming wave of unit amplitude plus a reflected wave and, far to the right, by a transmitted wave, whence,
\begin{equation}
c_n (k) = \left\{  \begin{array}  {l@{\quad \quad}l}
e^{in\theta_k} + re^{-in\theta_k}, & {\rm as} \quad n \rightarrow - \infty,\\
\tau e^{in \theta_k}, &  {\rm as} \quad n \rightarrow + \infty, 
\end{array} \right.
\label{eq31}
\end{equation}
where $r$ ($\tau$) is the reflection (transmission) amplitude.
The scattering potential for an electron propagating through the atomic wire with a dimer impurity is, in reduced notation,
\begin{equation}
V = 2\beta [ z_0 | 0 \rangle \langle 0 | + z_1 | 1 \rangle \langle 1 | + \gamma (|0 \rangle \langle 1| + |1 \rangle \langle 0|)] ,
\label{eq32}
\end{equation} 
where
\begin{equation}
z_j = (\alpha_j - \alpha)/ 2\beta ~,~ \gamma = (\beta_{01} -\beta)/ 2\beta ~,~ j=0,1,
\label{eq33}
\end{equation}
are the reduced dimer-site energies and the intra-dimer coupling, respectively.

For $n \rightarrow + \infty$, as in (\ref{eq31}), we can write (\ref{eq29}) in terms of the scattering-potential  matrix elements $V(l,m) = \langle l|V|m \rangle$ as 
\begin{equation}
\tau = 1 + (\omega \beta)^{-1} \sum_{l,m} t^{-l} V(l,m) c_m(k) ,
\label{eq34}
\end{equation}
by means of (\ref{eq30}) and where
\begin{equation}
\omega = -2i \sin \theta_k ~,~ t=e^{i \theta_k} ,
\label{eq35}
\end{equation}
with the $c_m(k)$'s in (\ref{eq34}) to be found.  On using (\ref{eq32}), we can express (\ref{eq34}) explicitly as
\begin{equation}
\tau = 1 + (\omega \beta)^{-1} [ V(0,0)c_0 + t^{-1} V(1,1) c_1 + V(0,1) c_1 + t^{-1} V(1,0) c_0 ],
\label{eq36}
\end{equation}
which by (\ref{eq32}) becomes
\begin{equation}
\tau = 1 + 2 \omega^{-1} [(z_0+ \gamma t^{-1} ) c_0 + (z_1 t^{-1} +\gamma)c_1].
\label{eq37}
\end{equation}

To find the coefficients $c_0$ and $c_1$ in (\ref{eq29}), we take $n=0$ and $n=1$, whereby we have 
\begin{equation}
c_0 = c_0^0 + (\omega \beta)^{-1} \sum_{l,m} t^{|l|} V(l,m) c_m ,
\label{eq38}
\end{equation}
\begin{equation}
c_1 = c_1^0 + (\omega \beta)^{-1} \sum_{l,m} t^{|1-l|} V(l,m) c_m ,
\label{eq39}
\end{equation}
On utilizing (\ref{eq32}) and (\ref{eq35}), these equations become
\begin{equation}
c_0 = 1 + 2 \omega^{-1} [(z_0+\gamma t) c_0 + (z_1+\gamma t^*) t c_1],
\label{eq40}
\end{equation}
\begin{equation}
c_1 = t + 2 \omega^{-1} [(z_0+\gamma t^*) t c_0 + (z_1+\gamma t) c_1],
\label{eq41}
\end{equation}
whose solutions are
\begin{equation}
c_0 = (1-2 z_1 t)/ \Delta ,
\label{eq44}
\end{equation}
\begin{equation}
c_1 = (1 + 2 \gamma) t / \Delta ,
\label{eq45}
\end{equation}
where
\begin{equation}
\Delta = 1 - 2 \omega^{-1} (P-2Qt),
\label{eq46}
\end{equation}
with
\begin{equation}
P = z_0+z_1  ~,~ Q = z_0 z_1 - \gamma -\gamma^2 .
\label{eq46a}
\end{equation}
Thus, on using (\ref{eq44}) to (\ref{eq46a}), equation (\ref{eq37}) provides the dimer transmission coefficient
\begin{equation}
\tau = (1 + 2 \gamma)/ \Delta ,
\label{eq47}
\end{equation}
whereby the transmission probability is given by
\begin{equation}
T = \tau \tau^* = |c_1|^2 = (1+2 \gamma)^2 / |\Delta|^2 ,
\label{eq48}
\end{equation}
which, by dint of (\ref{eq35}), (\ref{eq46}) and (\ref{eq46a}), can be written as
\begin{equation}
T = {{(1+2\gamma)^2 (4-X^2)} \over {(1-2Q)^2 (4-X^2) + 4(P-QX)^2}}  ,
\label{eq48a}
\end{equation}
with $X$ given by (\ref{eq12}). Thus, we see that $T \equiv T(E)$ in (\ref{eq48a}), since
$z_0$, $z_1$ and $\gamma$ in (\ref{eq33}) and (\ref{eq46a}) are all functions of energy $E$.

It is interesting to note that (\ref{eq47}) can be expressed as
\begin{equation}
\tau = c_1 t^{-1} , 
\label{eq49}
\end{equation}
via (\ref{eq45}), which, in turn, yields
\begin{equation}
c_1 = \tau t = \tau e^{i \theta_k} ,
\label{eq50}
\end{equation}
by (\ref{eq35}), which agrees with the form of (\ref{eq31}) for $n=1$.

Finally, we introduce the average value of the transmission probability
\begin{equation}
\bar{T} = (E_u-E_l)^{-1} \int_{E_l}^{E_u} T(E) dE ,
\label{eq51}
\end{equation}
as a measure of the overall transmittivity of a renormalized dimer, $E_l$ ($E_u$) being the lower (upper) energy band-edge of the leads.

\section{Results and Discussion}

Insight into the effects of quantum interference on the electronic transmission properties of substituted-benzene molecules is gained by investigating the fluctuating behaviour of the $T(E)$ curves of each of these molecules.  As we shall see, each $T(E)$ graph has unique distinguishing features, which identify it with its parent benzene molecule.

Although the transmission properties of the benzene molecule received a comprehensive treatment by Solomon, Hansen and their co-workers \cite{ref4}, a study via the renormalized-dimer approach is included in the present work for purposes of completeness and comparison with the transmission curves of each of the double- and triple-benzene molecules considered.

While the benzene $T(E)$ graph spreads its curve's profile across the lead's entire energy spectrum, between
the band-edges $E= \pm 1$, the actual electron transmission process takes place at $E=0$, the Fermi energy level.  The numerical values for single benzenes, reported in this section, are based on a separate analytical treatment of the $T(E)$ graphs, which are to be published elsewhere.  Here, we confine ourselves to a descriptive overview of the main details of the $T(E)$ graphs for each of the p-, m- and o-benzenes, as they pertain to the quantum interference effects. In the course of the underlying calculations, the parameter values chosen were the atomic-site energy $\alpha = 0$ and the bond energy $\beta = -0.5$.  These values apply to both the benzene molecules and their leads.

\subsection{Single-benzene circuits}

In Figure 7  is collected the $T(E)$ graphs of the p-benzene (solid curve), m-benzene (dashed curve) and o-benzene (dash-dotted curve) structures. Such a figure arrangement facilitates the comparison of the different spatial variations encountered in the benzene graphs. 
\begin{figure}[htbp]
\includegraphics[width=15cm]{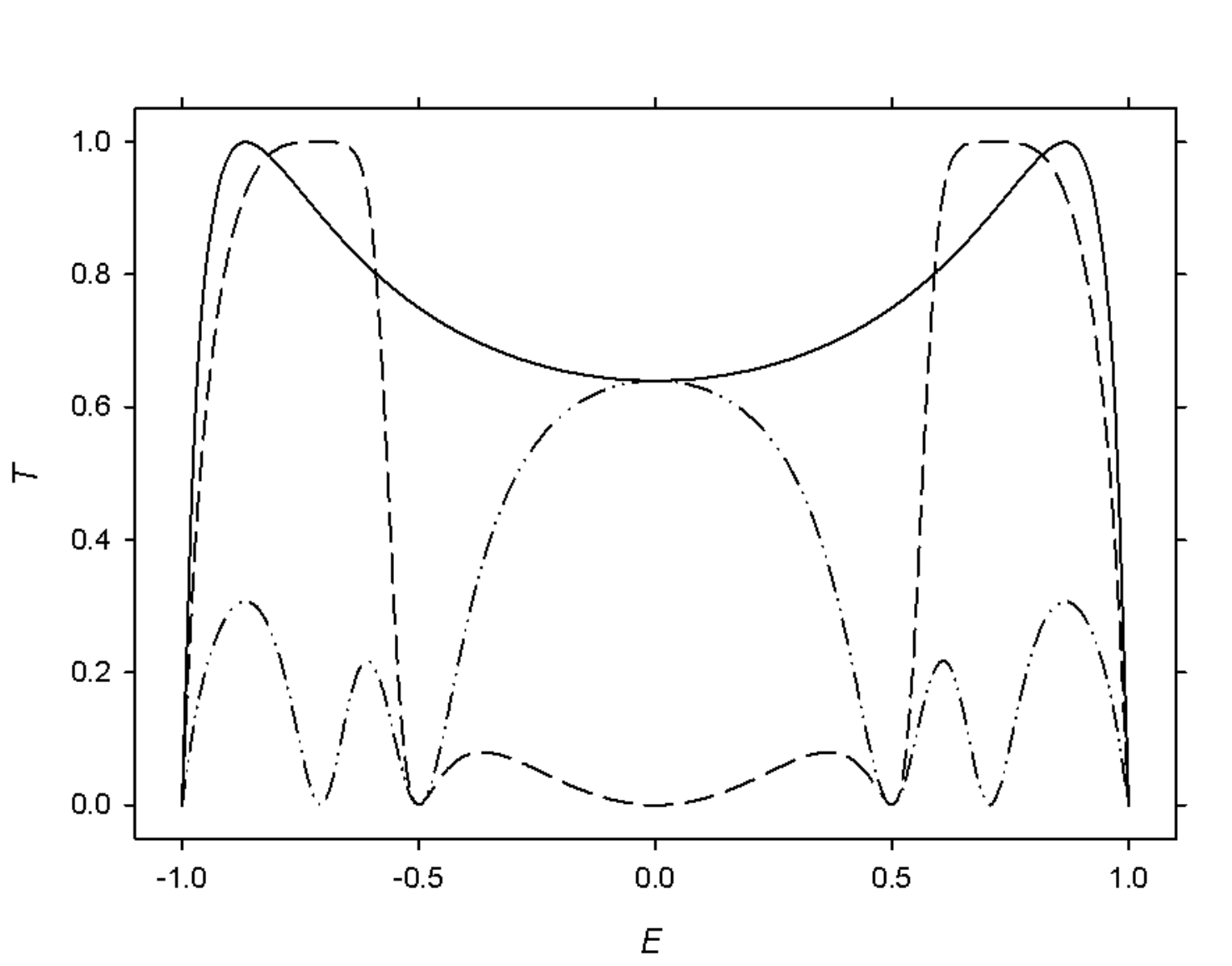}
\caption{Transmission $T$ versus energy $E$ for p- (solid curve), m- (dashed) and o-benzene (dash-dotted).}
\label{fig7}
\end{figure}

The p-benzene graph results from the constructive interference between  the two equal electronic pathways created by the atomic sites' (1,4) configuration.  This graph displays a curve of simple architecture with two resonances at $E= \pm 0.87$, separated by a ``high valley'' with a minimum of $T=0.64$ at $E=0$.  Here, we note that the electron transfer, which occurs at $E=0$, benefits from the high value of $T(0)$. Another feature is the anti-resonances located at $E= \pm 1$, the leads' band edges.  In view of the fact that $T(E)=T(-E)$, the graph possesses a mirror-image symmetry about $E=0$; such symmetry also occurs in the m- and o-benzene graphs.  

In the m-benzene graph, we witness the effect of destructive interference created by the unequal lengths of the electronic pathways in the (1,5) configuration of  the atomic sites. This type of interference completely restructures the p-benzene $T(E)$ profile and enriches it with interesting new features. These include widely separated slender resonance peaks at $E= \pm 0.71$ with anti-resonances at $E= \pm 0.5$ and at the usual limits of $E=\pm 1$.\footnote{Since the anti-resonances at the leads' band edges ($E=\pm 1$) have the property that $T(\pm 1)=0$, they are present in all the $T(E)$ spectra, and differ from the other class of anti-resonances, which lie between the band edges, and are those created by quantum interference.} Meanwhile, a small ``hill'' appears on either side of the antiresonance at $E=0$ at energies $E=\pm 0.37$ and whose height rises to $T=0.08$.  The removal ($T=0$) of the electron presence at the Fermi level makes the m-benzene the poorest electron transmitter of the three single benzenes.

Turning to the o-benzene $T(E)$ graphs, we are immediately aware of the broad central peak, whose apex on the $T$-axis surprisingly coincides tangentially with the bottom of the p-benzene minimum at $T=0.64$. Hence, the criterion that $T(0)>0$ is satisfied for electron transfer to occur across the Fermi level, even though the (1,5) configuration of the atomic sites produces unequal lengths of the electronic pathways, thus resulting in destructive interference.  This finding is in marked contrast to that of the m-benzene situation, where destructive interference was also present.  Moving on, we observe that the two resonance peaks at $E= \pm 0.71$ in the m-benzene graph have been replaced by two much smaller peaks, with coordinates at $E= \pm 0.61$ and $E= \pm 0.87$ and heights $T=0.22$ and $T=0.3$ respectively. These peaks lie inside the two m-benzene resonances, while between the peaks there are anti-resonances at $E= \pm 0.5$ and $\pm 0.7$, besides the two usual anti-resonances at $E=\pm 1$. Unlike the other two single-benzene $T(E)$ graphs, there are no resonances at all in the o-benzene case.

According to Table 1, where the average values $\bar{T}$ are compiled from equation (\ref{eq51}), the $\bar{T}$-value of 0.76 for p-benzene is approximately twice the $\bar{T}$-values for the m- and o-benzenes, which lie close together at 0.40 and 0.31, respectively. 
\begin{table}[htbp]
\renewcommand\arraystretch{1.2}
\begin{center}
\begin{tabular}{| l | l || l | l |}
\hline
\multicolumn{2}{|c|}{singles} & \multicolumn{2}{|c|}{$\bar{T}$}\\
\hline
\multicolumn{2}{|c|}{p} & \multicolumn{2}{|c|}{0.76}\\
\multicolumn{2}{|c|}{m} & \multicolumn{2}{|c|}{0.40}\\
\multicolumn{2}{|c|}{o} & \multicolumn{2}{|c|}{0.31}\\
\hline
\hline
series & $\bar{T}$ & parallel &  $\bar{T}$\\
\hline
pm & 0.35 & pm  & 0.32\\
po & 0.23 & po  & 0.25 \\
mo & 0.10 & mo & 0.37\\
pp & 0.65 & pp & 0.69\\
mm & 0.36 & mm & 0.26\\
oo & 0.17 & oo & 0.56\\
pmo & 0.13 & pmo & 0.12\\
pom & 0.10 & & \\
mpo & 0.09 &  & \\
ppp & 0.62 & ppp & 0.60\\
mmm & 0.34 & mmm & 0.17\\
ooo & 0.16 & ooo  & 0.47\\
ppm & 0.35 & ppm & 0.49\\
pmp & 0.36 &  &  \\
mmp & 0.33 & mmp & 0.27\\
mpm & 0.30 & & \\
ppo & 0.21 & ppo & 0.28\\
pop & 0.24 & & \\
oop & 0.16 & oop & 0.23\\
opo & 0.15 & & \\
mmo & 0.10 & mmo & 0.29\\
mom & 0.08 &  &\\
oom & 0.07 & oom & 0.35\\
omo & 0.12 & & \\
\hline
\end{tabular}
\end{center}
\caption{Values of $\bar{T}$ from equation (\ref{eq51}) for the various circuits. Note that parallel circuits,
such as pmo, pom and mpo, are indistinguishable, so their entries are not repeated.}
\label{tab1}
\end{table}
This $\bar{T}$-ordering clearly confirms the dominance of the p-benzene transmission \cite{ref4}, but fails to predict that the o-benzene transmission is, in fact, somewhat larger than the m-benzene one.  Such a situation arises because, in the averaging process, each energy level $E$ in the spectrum is considered, albeit weighted by its transmission probability $T(E)$. In doing so, it neglects the significance of the role played by the Fermi level in enhancing the $T(0)$-value of the electron transmission probability. Nevertheless, the value of $\bar{T}$ serves as one useful measure of transmission, because the averaging process considers the overall structure of the $T(E)$ curve.  For example, we note that $\bar{T}$ typically decreases with an increasing number of antiresonances because, as expected, they act to inhibit transmission.

\subsection{Series circuits}

The $T(E)$ graphs for two benzene molecules in series are shown in Figures 8 and 9.  The former figure illustrates circuits of the homogeneous type, while the latter one displays circuits of the heterogeneous type (noting that a circuit such as pm is indistinguishable from mp).  Turning first to Figure 8, we can see the effect of adding an extra molecule by comparison with the single-benzene cases of Figure 7. 
\begin{figure}[htbp]
\includegraphics[width=15cm]{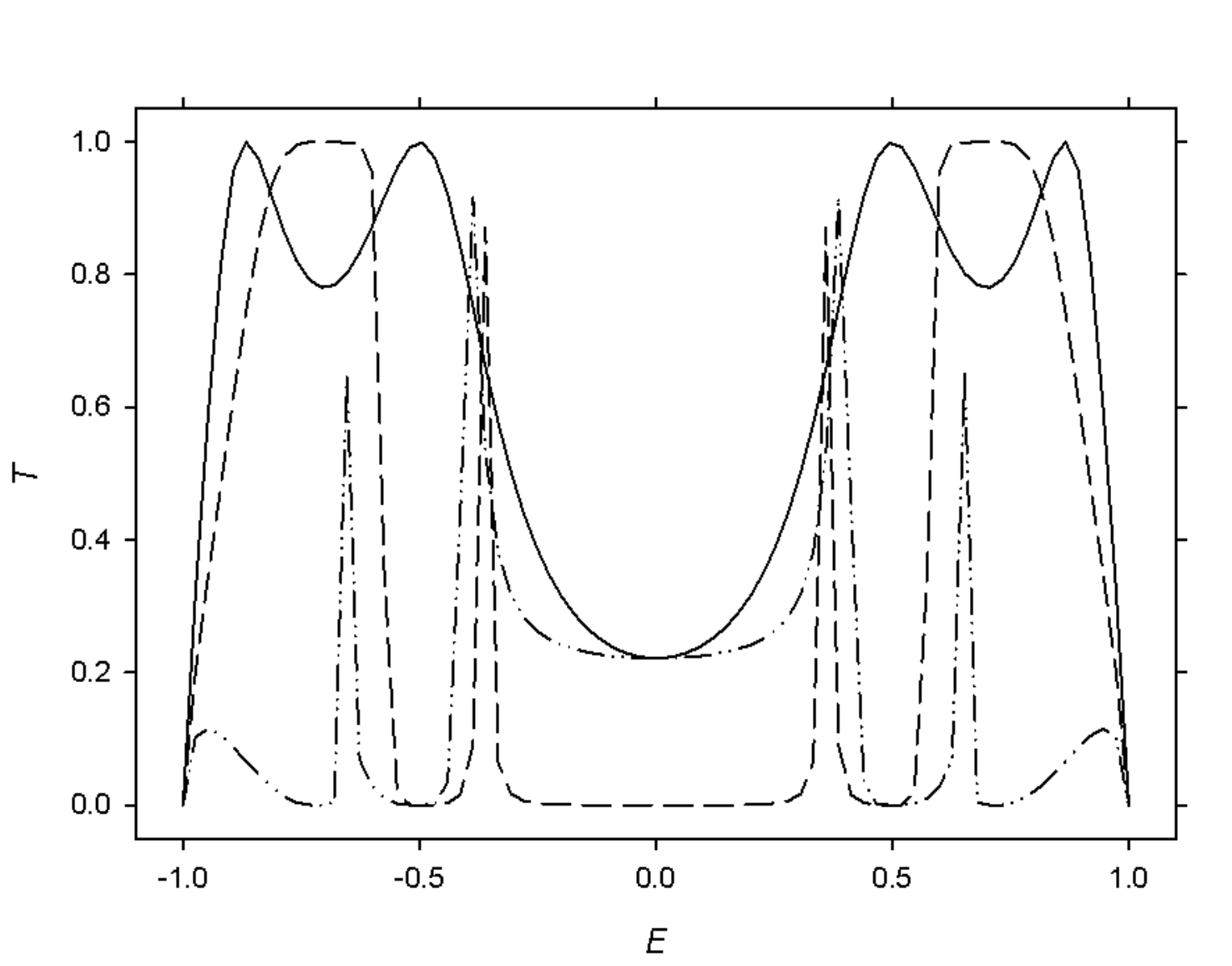}
\caption{Transmission $T$ versus energy $E$ for pp- (solid curve), mm- (dashed) and oo-benzene (dash-dotted) in series.}
\label{fig8}
\end{figure}
The curve for pp (solid curve) is an interesting contrast to that for p, the dominating features being the splitting of the p resonances at $E= \pm 0.87$ into two pairs at $E = \pm 0.87$ and $E= \pm 0.5$, which perhaps seems reminiscent of the splitting of degenerate states. While the appearance of extra resonances would seem to indicate better transmission properties for the double-benzene circuit, there is a counteracting effect due to the drastic lowering of the minimum at $E=0$, from $T=0.64$ in p-benzene down to $T=0.22$ in pp-benzene.  This lowering of the minimum at the Fermi level serves to inhibit the transmittivity of pp-benzene, with the result that it is a poorer transmitter than p-benzene. This point is reinforced by comparing the $\bar{T}$-value of 0.65 for pp with that of 0.76 for p, in Table 1. The situation for mm-benzene (Figure 8, dashed curve) is rather different.  We notice that the resonances at $E= \pm 0.71$ in m-benzene are maintained, while the smaller peaks at $E= \pm 0.37$ are actually heightened (and narrowed) to become resonances themselves.  However, all the anti-resonances, and in particular the one at $E=0$, persist, resulting in only marginal weakening of the poor transmission of m.  This is also evidenced by the $\bar{T}$-value of 0.40 for m dropping down only to 0.36 in mm.  Next we come to oo-benzene (Figure 8, dash-dotted curve), where again significant changes in the $T(E)$ curve arise, compared to the o-benzene case, resulting in a very different-looking curve. The outermost pair of peaks, at $E = \pm 0.87$, have been shifted somewhat to $E= \pm 0.95$, but diminished greatly, to heights of $T=0.12$, while the inner peaks at $E= \pm 0.61$ have been heightened into resonances, at about the same energy.  Meanwhile, the maximum, dominating the $T(E)$ curve for o, has been restructured into a minimum of $T=0.22$ at $E=0$, flanked by a pair of newly-created resonances at $E= \pm 0.40$.  This massive restructuring of the o curve into that for oo, and in particular the lowering of $T(0)$, results in a significant drop in the transmission properties of this already-poor transmitter.  As a further indicator, we notice that $\bar{T}$ drops from 0.31 for o down to 0.17 for oo.

Next, we consider the heterogeneous circuits shown in Figure 9. 
\begin{figure}[htbp]
\includegraphics[width=15cm]{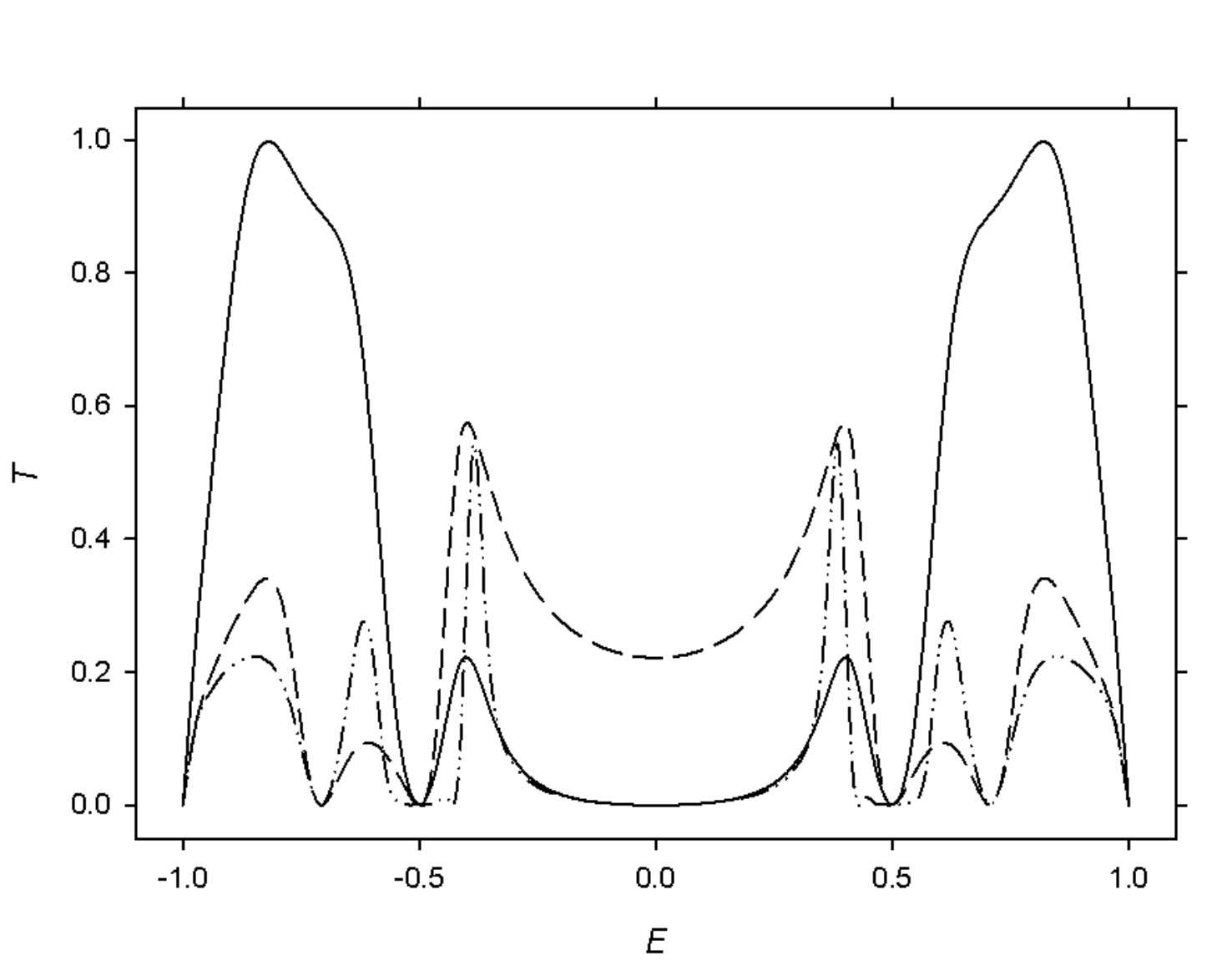}
\caption{Transmission $T$ versus energy $E$ for pm- (solid curve), po- (dashed) and mo-benzene (dash-dotted) in series.}
\label{fig9}
\end{figure}
 It is apparent that none of the three ``mixed'' curves bears a close resemblance to that of any individual single circuit, but instead seem to incorporate a mixture of features seen in the parents.  The graph of pm-benzene (solid curve) bears some resemblance to that of m, although the resonances, which are located at $E= \pm 0.81$, are situated between those for m and p.  Likewise, the smaller peaks, presumably originating with the m constituent, are shifted outwards to $E= \pm 0.4$, and also heightened due to the p.  The anti-resonance at $E=0$, seen in m, persists in pm and contributes to the circuit being a rather poor transmitter.  The po graph (dashed curve) is another blend of features. The two outer pairs of peaks, along with the anti-resonances, have been inherited from the o graph, while the inner pair of peaks, with a minimum of $T(E)$ between them, is more reminiscent of the p graph. The fact that $T(E)$ is low at virtually all values of $E$, and in particular near the Fermi level, results in this circuit being a rather poor transmitter.  Lastly, the graph for mo-benzene (dash-dotted curve) again shows the features of both parent curves. Most importantly, perhaps, is that every anti-resonance in m or o gives rise to an anti-resonance in mo and, in particular, at $E=0$.  With $T(E)$ being very low at all $E$, except near the two highest peaks ($E= \pm 0.38$), the result is a very poor transmitter, as is also evidenced by $\bar{T} = 0.1$.

We turn now to series circuits consisting of three benzene molecules, beginning with the three homogeneous types, shown in Figure 10.
\begin{figure}[htbp]
\includegraphics[width=15cm]{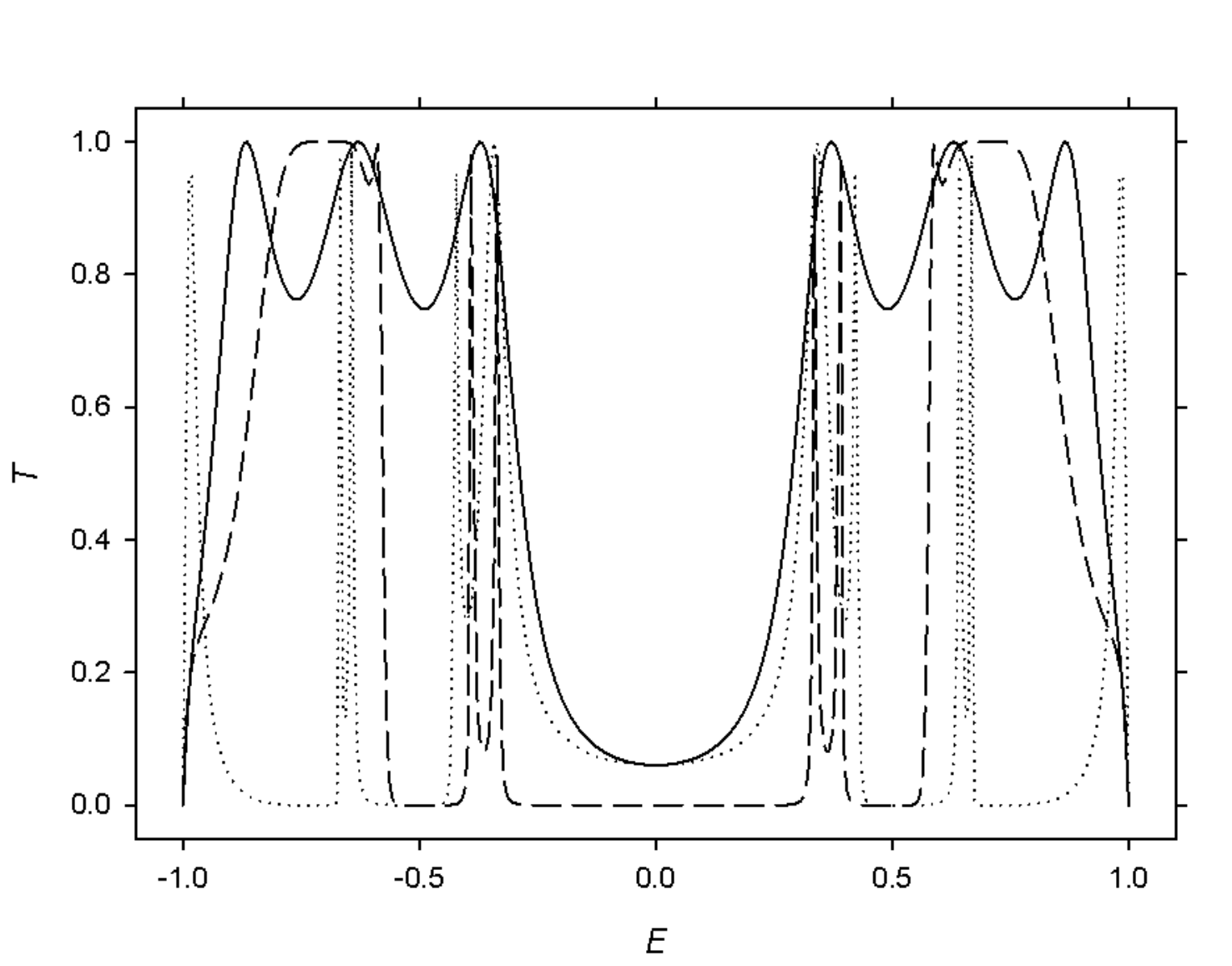}
\caption{Transmission $T$ versus energy $E$ for ppp- (solid curve), mmm- (dashed) and ooo-benzene (dotted) in series.}
\label{fig10}
\end{figure}
The graph for ppp (solid curve) is an interesting extension of that for p (Figure 7) and pp (Figure 8). In going from p to pp, we saw that an extra pair of resonances splits off from the original pair. In going now to ppp, a third pair of resonances arises. Although it might be supposed that 6 resonances would strengthen the overall transmissitivity of the circuit, this is counteracted by the very low value of $T(0) = 0.06$, with $\bar{T} = 0.62$ suggesting that ppp is a marginally weaker transmitter than pp, but with both being substantially weaker than p. Turning to mmm (dashed curve), we again observe the splitting of both pairs of resonances in the mm curve (Figure 8). All anti-resonances persist, including the key one at $E=0$, resulting in relatively unchanged transmittivity ($\bar{T} = 0.34$).  Similar behaviour appears in the last graph, that of ooo (dotted curve), where once again we see that each resonance of oo splits to create a pair of resonances in ooo, but additionally, the outermost peaks (at $E= \pm 0.985$) are heightened into two new resonances.  As always, all anti-resonances persist, but there is a substantial lowering of $T(0)$ to 0.06, resulting in a weak transmittivity.

Turning to heterogeneous triples, there is a total of 15 distinct possibilities, many with very similar graphs. Only the three graphs utilizing all three benzene types are shown in Figure 11. 
\begin{figure}[htbp]
\includegraphics[width=15cm]{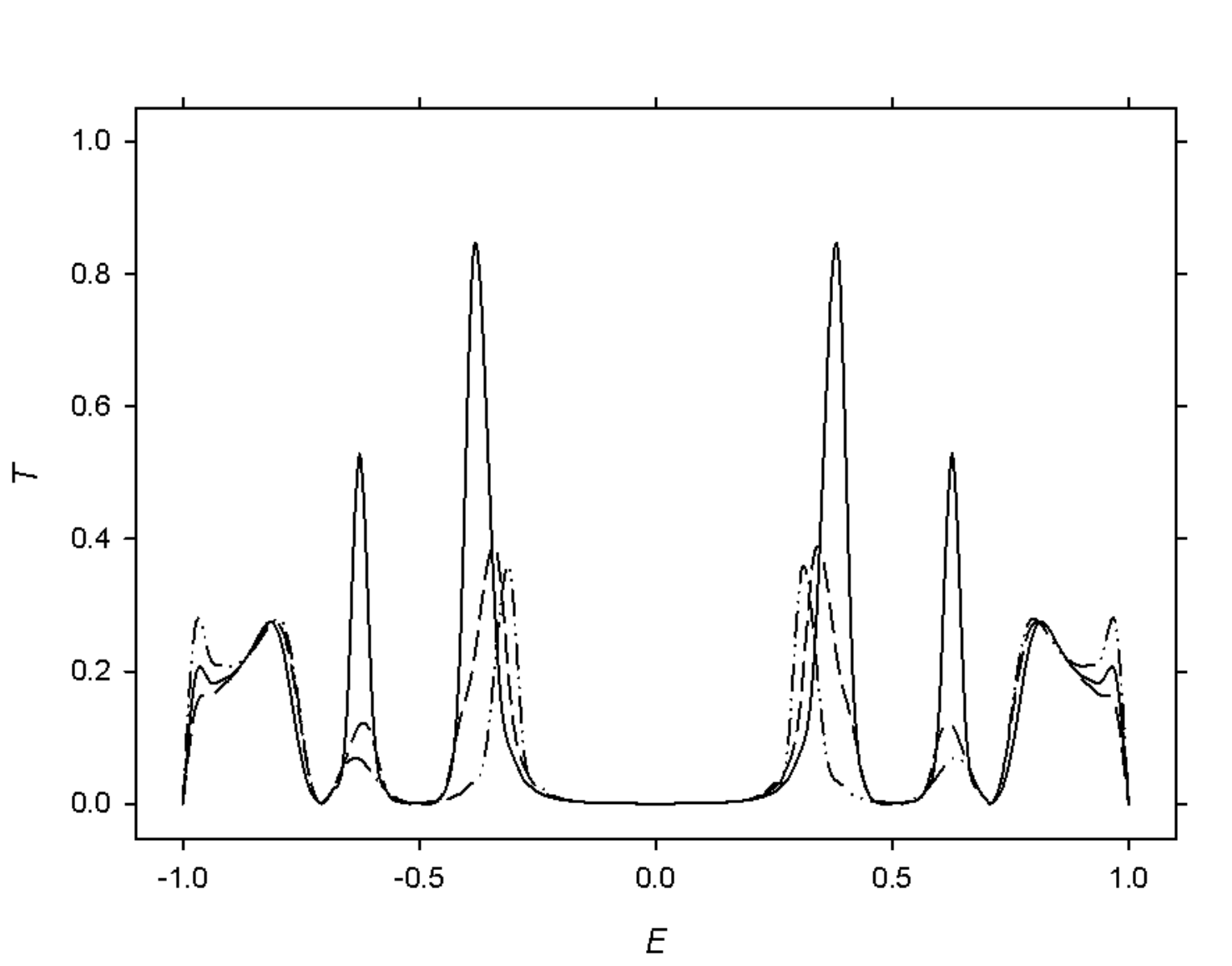}
\caption{Transmission $T$ versus energy $E$ for pmo- (solid curve), pom- (dashed) and mpo-benzene (dash-dotted) in series.}
\label{fig11}
\end{figure}
These three graphs are themselves quite similar, having the same anti-resonances (namely, those occurring in m or o), but no resonances, although the peaks are of varying (but typically low) heights.  With the anti-resonance at $E=0$ creating a broad spectrum of low transmission near the Fermi level, these three circuits are unsurprisingly all poor transmitters, with $\bar{T}$ around 0.1 for all of them (see Table 1). Of the other heterogeneous triples, none is a particularly good transmitter, as indicated in Table 1.  Using $\bar{T}$ as a measure, the best transmitters tend to be those involving p and avoiding o.

We conclude this subsection with some comments about the general trends in the series circuits. It is noted that all graphs are symmetric, $T(E) = T(-E)$, and that anti-resonances arise only from an anti-resonance in a single benzene (specifically, m or o).  As seen from the table of $\bar{T}$-values, a general concept is that longer circuits typicially provide poorer transmittivity. The addition of a p anywhere in a series creates no new anti-resonances, and tends to be modestly suppressive of the transmission. The effect of adding an m to a series is to create anti-resonances at $E=0,~ \pm 0.5$, if not already present.  There typically appears to be some enhancement of $T(E)$ near the band edges, but, more significantly, suppression of it near the band center.  The effect of adding an o to a series is to create anti-resonances at $E= \pm 0.5, ~\pm 0.7$, again if not already present.  There is often some enhancement of $T(E)$ near the band center, but overall it is very suppressive of transmission. The biggest effect typically arises from lengthening the circuit from one benzene ring to two, while adding a third ring has a lesser impact.

\subsection{Parallel circuits}

The $T(E)$ graphs for two benzene molecules in parallel are shown in Figure 12, for the homogeneous cases, and in Figure 13, for the heterogeneous ones.
\begin{figure}[htbp]
\includegraphics[width=15cm]{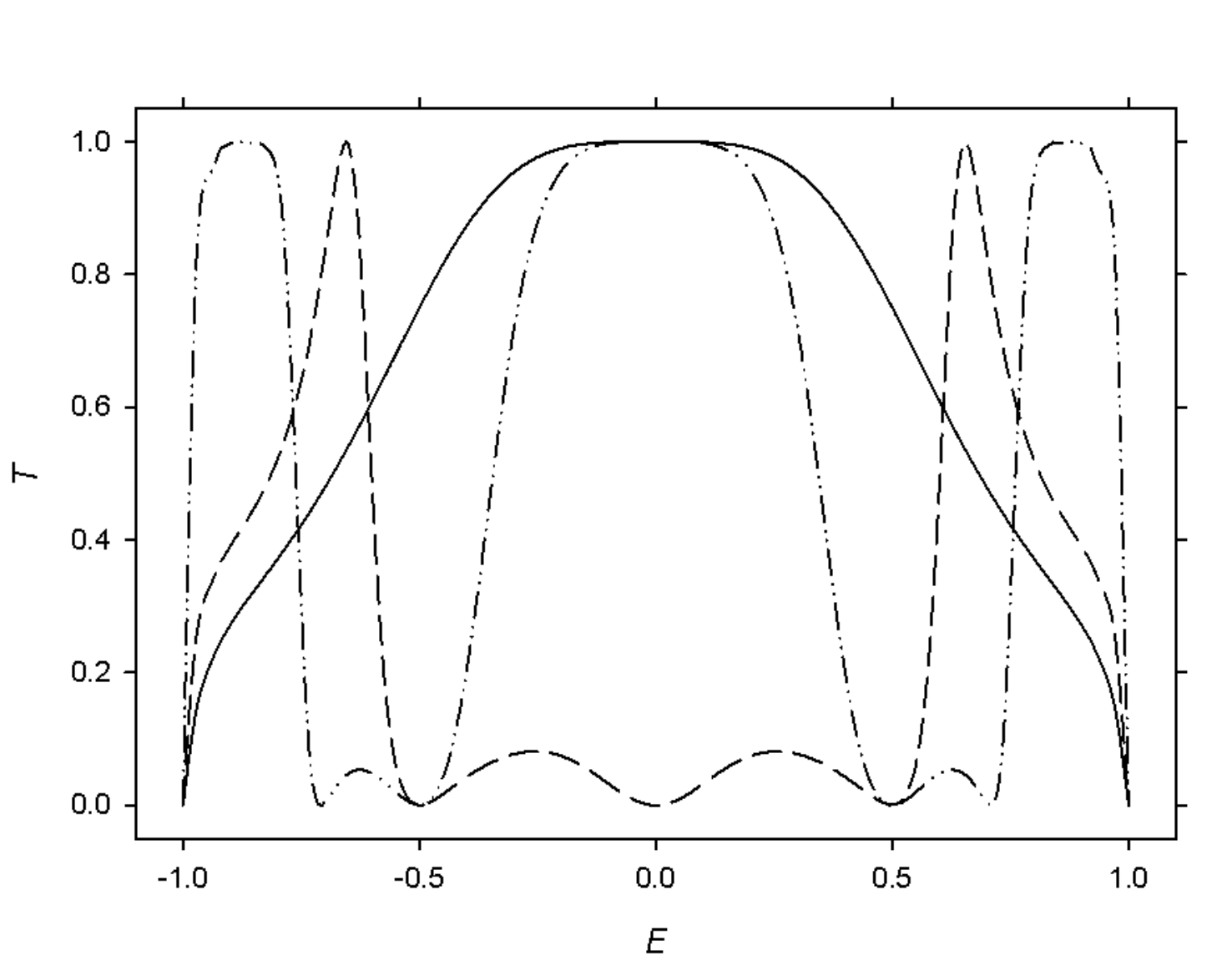}
\caption{Transmission $T$ versus energy $E$ for pp- (solid curve), mm- (dashed) and oo-benzene (dash-dotted) in parallel.}
\label{fig12}
\end{figure}
Turning first to the former, we examine their similarities to, and differences from, their single benzene counterparts.  The (solid) curve for pp bears little resemblance to its p parent (Figure 7), possessing just a single, albeit very broad, resonance peak at $E=0$, but no anti-resonances, suggesting a very good transmitter, which is reinforced from Table 1 by the fact that $\bar{T} = 0.69$.  By contrast, the graph for mm (dashed curve) shows an obvious similarity to that of m, reproducing the energies of all of its anti-resonances and both resonances.  In particular, with an anti-resonance at $E=0$ and low $T(E)$ at surrounding energies, mm is as poor (or poorer) a transmitter as m. Lastly, we come to oo (dash-dotted curve), where there is a strong resemblance to the o curve, due to all the anti-resonances being the same (as is, in fact, always the case for homogeneous parallel circuits). However, the important difference between the two curves is that three of the peaks in the o graph, including the largest one centered at $E=0$, are now heightened into resonances, which has the effect of greatly improving the transmission (compare $\bar{T} = 0.31$ for o to $\bar{T} = 0.56$ for oo).  This fascinating feature illustrates the novel possibilities that can arise by experimenting with different molecular combinations. Comparing Figure 12 to the corresponding double-benzene circuits in Figure 8, it is immediately noticeable that the parallel situation is strikingly different from the series one.  Despite the fact that each parallel circuit possesses the same anti-resonances (or lack thereof) as its series counterpart, the $T(E)$ curves have few, if any, features in common, such as number and position of resonances, heights of peaks, locations of relative minima, etc. These dissimilarities emphasize the point that series versus parallel circuits are very different constructions.

Looking next at the heterogeneous double-benzene parallel circuits (Figure 13), the most striking feature is that two of the three curves (namely, those for pm and mo) are asymmetric about $E=0$, i.e., $T(E) \ne T(-E)$. 
\begin{figure}[htbp]
\includegraphics[width=15cm]{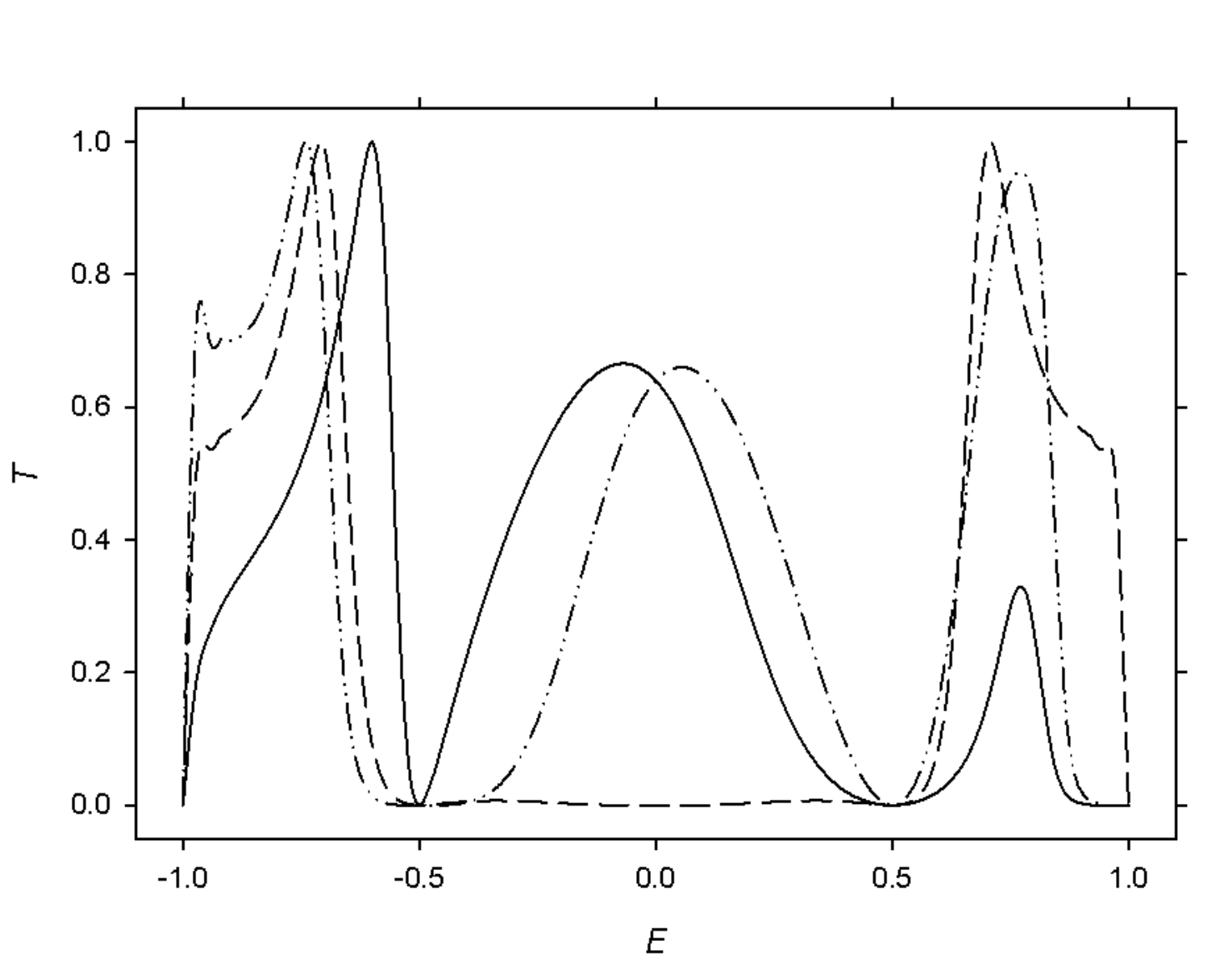}
\caption{Transmission $T$ versus energy $E$ for pm- (solid curve), po- (dashed) and mo-benzene (dash-dotted) in parallel.}
\label{fig13}
\end{figure}
As it turns out, this asymmetry occurs only when at least one m-benzene is placed in parallel with one (or more) p- or o-benzenes. This feature will be explained in more detail in the following paragraph. Closer examination of the curves indicates that none of them bear a close resemblance to those for the single-benzenes, from which they are constructed, in terms of numbers and positions of resonances and anti-resonances, etc. Of the three curves, that for po (dashed curve) clearly has the lowest overall T(E) values, especially in the crucial region around the Fermi level, and thus can be expected to be the poorest transmitter (as is also indicated by comparing values of $\bar{T}$ in Table 1), although none show strong transmission properties.

The asymmetry exhibited by the pm and mo curves in Figure 13 originates in the parity properties of the rescaled $\beta$'s, arising in the renormalization procedure, and how that affects the transmission probability $T(E)$ in (\ref{eq48a}). In looking at the rescaled $\beta$'s for the single dimers, it is obvious that $\bar{\beta}_p$ in (\ref{eq11}) and $\bar{\beta}_o$ in (\ref{eq16}) are even, while for the m dimer, $\bar{\beta}_m$ in (\ref{eq14}) is odd. In a series circuit, replacing $E$ by $-E$ (or equivalently, $X$ by $-X$) amounts to, at most, a change in sign, but not in magnitude, in  $\tilde{\beta}_{16}$ (\ref{eq23})  for the renormalized dimer.  Consequently, $T(E)$ is unchanged and is thus symmetric with respect to $E$.  However, in a parallel circuit, the additive nature of $\tilde{\beta}_{78}$ in (\ref{eq28}) means that a sign change in some, but not all, of the $\beta_n$ results in a change in magnitude of $\tilde{\beta}_{78}$, leading to a change of value of $T$, so that $T(E) \ne T(-E)$, in general.  This situation occurs only when an m-benzene is in parallel with a p or an o, producing this interesting, and perhaps useful, asymmetry.

Finally, we arrive at three benzenes in parallel, for which there is a total of 10 possible distinct combinations. As a representative sampling, the curves for the three homogeneous cases, plus that for pmo, are shown in Figure 14. 
\begin{figure}[htbp]
\includegraphics[width=15cm]{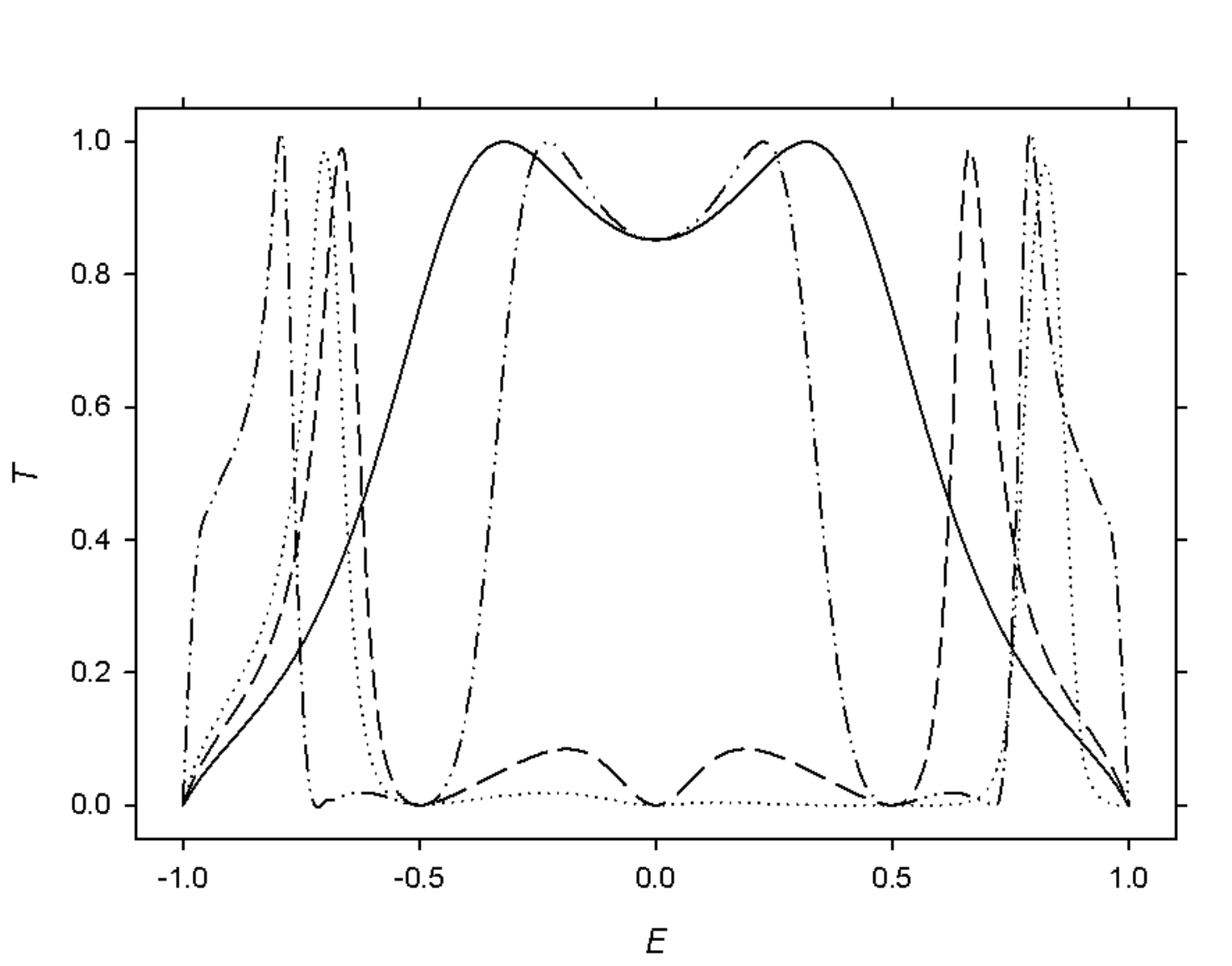}
\caption{Transmission $T$ versus energy $E$ for ppp- (solid curve), mmm- (dashed), ooo- (dash-dotted) and pmo-benzene (dotted) in parallel.}
\label{fig14}
\end{figure}
The graph for ppp (solid curve) exhibits twin resonances at $E= \pm 0.4$, with a high minimum value of $T(0)=0.85$ between them, indicating that it should be a good transmitter.  The graph is quite similar to that for pp in parallel (Figure 12), suggesting a similar transmittivity, albeit somewhat weaker due to the minimum at $E=0$. The graph for mmm (dashed curve) is also similar to the corresponding one for mm (Figure 12), but with the resonances narrower. That feature along with the anti-resonance at $E=0$ implies poor transmission. The graph for ooo (dash-dotted curve) also resembles that for oo in parallel (Figure 12), but with the outer resonances narrowed, and the resonance at $E=0$ split into two resonances at $E= \pm 0.23$, with a high local minimum of $T(0)=0.85$ between them. This indicates that ooo is a reasonably good transmitter, albeit weaker than oo.  Lastly, the graph for pmo (dotted curve) is asymmetric (for the reason explained in the previous paragraph), but not particularly resembling that of any component. It has a resonance at $E \approx -0.7$, a near-resonance at $E \approx 0.8$ and several anti-resonances (most notably at $E=0$), with very low $T(E)$ across a wide range of $E$-values close to the Fermi level.  Consequently, one expects this circuit to have very low transmittivity. These inferences from the graphs of Figure 14 are reinforced by looking at the corresponding $\bar{T}$-values in Table 1. Of the remaining possible circuits (all of type xxy), whose graphs are not shown, these tend to resemble the heterogeneous xy parent, with the addition of an extra benzene ring often enhancing the $T(E)$ curve, especially near $E=0$.  In particular, the addition of a p to a pm or po circuit noticeably increases the transmittivity (see Table 1).

To conclude this subsection, we note some general trends among the parallel circuits. The homogeneous circuits, of the type x, xx and xxx, tend to have very similar $T(E)$ graphs, in which, for example, the anti-resonances of the single benzene persist into the more complicated circuits. By contrast, the heterogeneous circuits have graphs that generally do not resemble those for the corresponding singles. In particular, anti-resonances, resonances and maxima/minima of the single may not survive when another, different ring is added to the circuit. An overall trend is that larger circuits possess lower transmittivity, although there are exceptions.

\section{Conclusion}

Since a single benzene ring acts as a basic building block in the fabrication of more complex molecular structures, it is essential to gain a clear understanding of its electronic properties. In this way, a sound basis is established for investigating the transport aspects of the subsequent molecular circuits. Recently, a detailed molecular Green function analysis was performed of the electron transmission-energy spectra of the p-, m- and o-benzenes \cite{ref4}, where the resonance broadening, arising from the presence of the attached leads, was described in terms of first-order perturbation theory. In contrast, the present work adopts the renormalization-decimation approach, whereby the substituted-benzene in question is reduced to a corresponding dimer between its two leads. This complete system is then subjected to the Lippmann-Schwinger scattering technique, which provides the transmission versus energy function $T(E)$, whose form reflects the electronic structure of the original benzene molecule. The main advantage of the latter method lies in its decimation process, which enables the larger and more complicated molecular systems to be addressed more easily, as witnessed by the treatments of the benzene molecules in series and parallel circuits, whose analytical findings were fully discussed.  Moreover, we note that the resonance broadening now occurs without the need to invoke perturbation theory.

The calculated results show an abundance of interesting features. At the core of understanding these systems lies the fact that p-benzene is, by far, the strongest transmitter of the three types. A general, although not universal, aspect is that larger circuits usually have poorer transmission.The $T(E)$ graphs are typically symmetrical, except when an m-benzene is in parallel with p or o. The transmission curves are dominated by the numbers and positions of antiresonances and resonances (and other maxima). The results do contain some surprises, such as the transmission of o-benzene being increased by having 2 or 3 such molecules in parallel, which does not happen for p or m. Thus, the richness of these systems bodes well for future investigation and innovation.

\section{Keywords}

Benzene, molecular electronics, renormalization-decimation technique, semi-empirical calculations

\end{document}